\documentstyle[twocolumn,aps,prl,epsf,floats]{revtex}
\def\lsi{\raise0.3ex
\hbox{$<$\kern-0.75em\raise-1.1ex\hbox{$\sim$}}}
\def\gsi{\raise0.3ex
\hbox{$>$\kern-0.75em\raise-1.1ex\hbox{$\sim$}}}

\begin{document}
\twocolumn[\hsize\textwidth\columnwidth\hsize\csname
@twocolumnfalse\endcsname

\title{Homogeneous magnetic fields in fully anisotropic string cosmological 
backgrounds}
\author{ Massimo Giovannini}
\address{
{\it Institute for Theoretical Physics, University of Lausanne, 
BSP-Dorigny, CH-1015 Switzerland}}

\maketitle

\begin{abstract}
\noindent
We present new solutions of the string cosmological effective action 
in the presence of a homogeneous Maxwell field with pure magnetic component.
Exact solutions are derived in the case of space-independent
dilaton and vanishing torsion background. In our examples the four 
dimensional metric is either of Bianchi-type III and VI$_{-1}$ or 
Kantowski-Sachs.  
\end{abstract}

\vskip1.5pc]

The solutions of the low-energy string cosmological effective action
are, by their nature anisotropic \cite{pbb}.
 In this context, it is important to analyze
the role of the different possible sources of anisotropy in the low energy 
string effective action. An interesting and motivated 
source of anisotropy is represented by pure magnetic fields \cite{1}. 
Exact string cosmological solutions (containing a magnetic field)
were recently found for Bianchi-type I background 
geometries in the case of a vanishing torsion background 
and for a space-independent dilaton field \cite{1}. It was shown that
while the anisotropy can increase for some time, the string tension corrections 
(combined with the post big-bang evolution) are likely to force 
the anisotropy to decay.

The purpose of our paper is very simple: we want to generalize the 
Bianchi-type I solutions of the string cosmological effective action to the 
case where the homogeneous magnetic field is contained in more general 
homogeneous (but still anisotropic) metrics. In the 
context of the low-energy string effective action, the analysis of  
Bianchi-type classes has been performed (in the absence 
of any electromagnetic background) by Gasperini and Ricci \cite{2}.
 
In the Bianchi-type I case the homogeneous magnetic (or electric)
background is not constrained by the group theoretical properties of the
(Abelian) isometry group. In spite of this observation the solutions 
of the string cosmological effective action are not trivial 
in the sense that they cannot be found in the usual general relativity 
literature. The reason is that the dilaton field $\phi$ couples directly 
to the kinetic term of the Abelian gauge field. In four space-time 
dimensions 
such a coupling can be expressed as $e^{-\phi} F_{\mu\nu} F^{\mu\nu}$ 
where $\mu, \nu =0,...3$ and $F_{\mu\nu}$ is the Abelian gauge 
field strength. 
Bianchi models (different from Bianchi-type I) are characterized by 
a non-Abelian isometry group: the Killing vectors (leaving invariant 
the three-dimensional spatial submanifold) do not commute. 
 Therefore, we face the problem of 
accommodating an Abelian gauge field in a geometry whose algebraic 
structure is non-Abelian. The consequence of this remark is that  
{\em not all} non-Abelian isometry groups will be able to support 
a homogeneous electromagnetic field. More precisely, only Bianchi-type
I, II, III, VI$_{-1}$ and VII$_{0}$ can accommodate a homogeneous 
magnetic field polarized along a fixed direction \cite{3}. This 
observation holds in the case of general relativity (where the 
coupling of the metric to the Maxwell fields is dictated by the equivalence 
principle) and also in our case {\em provided} the dilaton field is 
space-independent and provided we deal with pure magnetic field
(i.e. the homogeneous components of the electric fields are all 
vanishing).

In order to show this point in detail let us look at the relevant 
Maxwell's equations. Then we will specify the background geometry
which is uniquely characterized by the group theoretical 
properties of the algebra of the Killing vectors. Finally, we will 
be able to show which Bianchi-type metrics can (or cannot) be compatible 
with the presence of a homogeneous (Abelian) magnetic field.
Let us start from the Maxwell's equations and from the 
related Bianchi identities
\begin{equation}
\partial_{\mu} \biggl(e^{- \phi} \sqrt{- g} F^{\mu\nu} \biggr) =0,\,\,\,\,
\partial_{\mu}\biggl(\sqrt{- g} \tilde{F}^{\mu\nu} \biggr) =0, 
\label{mx}
\end{equation}
where $\tilde{F}^{\mu\nu}$ is the dual field strength.. 

In the absence of electric fields and for the case of 
space-independent dilaton background the gauge field strengths 
can be written, in the language of differential forms, as 
\begin{equation}
F= \frac{1}{2} B_{i} \epsilon_{i j k} \sigma^{j} \wedge \sigma^{k},\,\,\,
^{\ast} F = - B_{i} \sigma^{i}\wedge\sigma^{0},
\label{2form}
\end{equation}
where $\sigma^{0}= dt$ and $\sigma^{i} = R^{i}_{j}\omega^{j}$ form the Cartan
 basis of one-form; the two-form $F$ represents the electromagnetic 
field and $^* F$ its dual. The matrices $R_{i}^{j}$ depend only on time. If 
we take, for instance Cartesian coordinates as local basis the relation 
between the exterior derivative of the invariant basis is given by :
\begin{equation}
d\omega^i = \frac{1}{2} C^{~~~i}_{j k} \omega^{j}\wedge\omega^{k}
\end{equation}
where $C_{j k}^{~~~i} $ are the structure constants of the three dimesional 
isometry group which appear in the commutation relations of the Killing 
vectors. For the case of Bianchi-type I we have that  
$C_{j k}^{~~~i} $ are identically 
zero (i.e. the isometry group is Abelian). For the other Bianchi-type
geometries the structure constants  are not all vanishing (i.e. the 
isometry group is non-Abelian).

By now expressing Eqs. (\ref{mx}) in the invariant basis of the $\omega^{i}$ 
and by recalling that the dilaton field is only time dependent 
we obtain a system of three linear equations, namely
\begin{eqnarray}
&& \dot{Z}_{a} =0,\,\,\, M_{j}C_{a b}^{~~~j}\epsilon_{c a b}=0,
\label{a1}\\
&& Z_{j} C_{a j}^{~~~j}=0,
\label{a2}
\end{eqnarray}
where the over-dot denotes differentiation with respect to $t$ and where
\begin{eqnarray}
&&
Z_{a} = \frac{1}{2}B_{i} \epsilon_{i j k} \epsilon_{a b c}R^{j}_{b} R^{k}_{c}, 
\nonumber\\
&& M_{j}= - \frac{1}{2}R^{i}_{j} B_{i}. 
\end{eqnarray}
The solution of the first of Eqs. (\ref{a1}) gives $Z_{a} = q_{a}$ where 
$q_{a}$ are constant vectors. Therefore Eqs. (\ref{a1})--(\ref{a2}) can 
be written as 
\begin{equation}
Z_{a}= q_a,\,\,\, q_a C_{j a}^{~~~j}=0,\,\,\,\, M_{j} C_{a b}^{j}= 0.
\label{a3}
\end{equation}
It is clear from these expressions that in the case where 
all the $C_{j k}^{~~~i}=0$ the magnetic field components 
are not constrained since all the equations (\ref{a3}) are 
identically statisfied for any vector $B_{i}$ and $q_a$.
In the Bianchi-type II case we have that the only non-vanishing 
structure constant is $C_{2 3}^{~~~1}=1$. In this case 
the first two equations of (\ref{a3}) are always satisfied, whereas
the third of Eqs. (\ref{a3}) constrains the magnetic field to vanish 
in one specific direction (i.e. the magnetic field can only have 
two independent components). Similarly, in the Bianchi-type III case
the only non-vansihing structure constant is $C_{2 3}^{2}=1$ and therefore 
both, the second and the third equation of (\ref{a3}) will 
be non trivially satisfied. Thus in Bianchi-type III models the 
magnetic field will have only one independent component: the other two 
will  have to vanish for the compatibility with the non-Abelian 
structure of the isometry group. 
In the Bianchi-type IV and V cases the structure constants are, 
respectively 
\begin{eqnarray}
&&C_{1 3}^{~~~1} = C_{2 3}^{~~~1} = C_{2 3}^{~~~2} = 1, \,\,\, {\rm IV},
\nonumber\\
&& C_{1 3}^{~~~1}= C_{2 3}^{~~~2} =1, \,\,\,\,\,\,\,\,\,\,\,\,{\rm V}.
\end{eqnarray}
From the second of Eq. (\ref{a3}) we see that, in the Bianchi-type 
IV and V, $q_3=0$. Moreover form the third of Eq. (\ref{a3}) we see 
that $B_k=0$. Therefore, in the Bianchi-type IV and V 
the isometry group is only compatible with a vanishing (pure) magnetic field.

In the Bianchi-type VI and VII the structure constants are, respectively, 
\begin{eqnarray}
&& C_{13}^{~~~1} = 1,\,\,\,\, C_{2 3}^{~~~2} =h,\,\,\,\,\,\,\,\,{\rm VI },
\nonumber\\
&& C_{3 2}^{~~~1} = C_{1 3}^{~~~2} =1 ,\,\,\,\, C_{2 3}^{~~~2} = h,\,\,\,{\rm VII}.
\end{eqnarray}
In the case of Bianchi-type VI a magnetic field can be 
only present (though with only one independent 
component) if $h=-1$ . Only in this case one can 
consistently satisfy the second of Eqs. (\ref{a3}).
In the Bianchi-type  VII geometry, for the same 
reason, a magnetic field (with one independent component)
can be present only if $h=0$.
Finally in the Bianchi-type VIII and IX metrics the structure 
constants are, respectively 
\begin{eqnarray}
&& C_{2 3}^{~~~1} = C_{1 2}^{~~~3} = C_{1 3}^{~~~2}=1,\,\,\,\, {\rm VIII},
\nonumber\\
&& C_{i j }^{~~~k} = \epsilon_{i j k},\,\,\,\,\,\,\,\,\,\,\,\,\,{\rm IX}.
\end{eqnarray}
In these two last cases we can see that the third equation
of (\ref{a3}) implies that $B_i R^{i}{j}=0$ which is 
identically satisfied only {\em in the absence of any magnetic 
background}. 

In conclusion we showed that 
{\em if the dilaton field only depends upon time} 
and {\em if the electric field is absent }
only the Bianchi-type I, II, III, VI$_{-1}$, VII$_{0}$ 
{\em are compatible with the presence of a magnetic 
background}. In the Bianchi-type I case {\em no constraint} 
on the components of the magnetic field is present. 
In the other Bianchi-type models the magnetic field 
{\em is always constrained}. In the Bianchi-type 
IV, V, VI$_{h\neq 1}$, VII$_{h\neq 0}$, VIII, IX cases
we cannot accommodate a pure (Abelian) magnetic field.

What we discussed, up to now is only {\em a sufficient 
condition} in order to accommodate a magnetic background 
with time dependent dilaton in a Bianchi-type geometry.
Our analysis does not guarantee the existence of 
explicit (exact) solutions.

 In the following, as an example,
 we want to present 
few exact solutions containing both a magnetic field, a time-dependent 
dilaton and a background geometry with non-Abelian 
isometry group.  
The action \footnote{We work in string units and we set the string 
tension $\alpha'$ equal to one} we ought to study is 
\begin{equation}
S= - \int d^4 x \sqrt{- g} \,e^{- \phi} \biggl[ R + 
g^{\alpha\beta} \partial_{\alpha}\phi 
\partial_{\beta}\phi + \frac{1}{4} F_{\mu\nu}F^{\mu\nu} \biggr]
\end{equation}
whose associated equations of motion can be written as 
\begin{eqnarray}
&&R - g^{\alpha\beta} \partial_{\alpha}\phi \partial_{\beta}\phi + 
2 g^{\alpha\beta} \nabla_{\alpha} \nabla_{\beta} \phi = 
- \frac{ 1}{4} F_{\alpha\beta} F^{\alpha\beta}, 
\label{phi}\\
&&R_{\mu}^{\nu} + \nabla_{\mu}\nabla^{\nu} \phi 
+ \frac{1}{2} F_{\mu\alpha}F^{\nu\alpha} =0,
\label{R}\\
&&\nabla_{\mu}\biggl[ e^{-\phi} F^{\mu\nu}\biggr] =0.
\label{F}
\end{eqnarray}
where $\nabla_{\mu}$ denotes covariant differentiation with respect 
to the background metric. Notice that in writing the previous action we 
assumed zero central charge deficit. This means that the six 
internal dimensions are assumed to be compactified with constant 
radius.
For sake of simplicity we choose to parametrize our line element as 
\begin{equation}
ds^2 = dt^2 - a^2(t) dx^2 - b^2(t) e^{ 2 \lambda x} dy^2 - c^2(t) dz^2.
\label{bianchi}
\end{equation}
Notice that for $\lambda = -1$ we have the Bianchi III line element, whereas 
the case $\lambda = -2$ we get the Bianchi VI$_{-1}$ case. 
In the metric of Eq. (\ref{bianchi}) the components of Eqs. 
(\ref{phi})--(\ref{F}) can be found by direct projection on 
the spatial vielbein. Care must be taken, however, in solving the 
generalized Maxwell equation (\ref{F}). Eq. (\ref{F}) 
together with the Bianchi identities imply, in the absence of sources, 
that a pure magnetic field can 
be accommodated along the $z$ axis so that $F_{\alpha\beta}F^{\alpha\beta} = 
2 B^2 /a^2 b^2$ ($B$ is a constant). With these specifications and by defining 
 the shifted time \footnote{ The over-dot denotes derivation with respect to the 
cosmic time coordinate $t$.} derivative of the dilaton  
$\dot{\overline{\phi}} = \dot{\phi} - (H+ F + G)$ together with the 
expansion rates along the three different spatial directions (i.e. $H= \dot{a}/a$, 
$F= \dot{b}/b$ and $G= \dot{c}/c$) we get to
\begin{eqnarray}
&&2 \ddot{\overline{\phi}} - \dot{\overline{\phi}}^2 - 
(H^2 + F^2 + G^2) + 2 \frac{\lambda^2}{a^2} + \frac{B^2}{2 a^2 b^2} =0,
\label{I}\\
&&  \ddot{\overline{\phi}}  - (H^2 + F^2 + G^2) =0,
\label{II}\\
&& H \dot{\overline{\phi}} - \dot{H} + \frac{\lambda^2}{a^2} 
+ \frac{B^2}{2 a^2 b^2} =0,
\label{III}\\
&&F\dot{\overline{\phi}} - \dot{F} + \frac{\lambda^2}{a^2} 
+ \frac{B^2}{ 2 a^2 b^2} =0,
\label{IV}\\
&&- \frac{\lambda}{a^2}( H - F) =0,
\label{V}
\end{eqnarray}
In order to solve exactly the previous system we can define a new time 
coordinate, namely, $dt= e^{- \overline{\phi} } d \eta$. By denoting 
with a prime the derivatives with respect to $\eta$ we can re-write 
the system of equations (\ref{I})--(\ref{V}) as 
\begin{eqnarray}
&&{\cal G}' = 0, \,\,\,\,\, {\cal H} = {\cal F}\,\,, 
\label{A}\\
&& {\cal H}' = \lambda^2 a^2 c^2  e^{-2 \phi} + \frac{B^2}{2} c^2 
e^{ - 2 \phi}\,\, ,
\label{B}\\
&&\phi'' + {\phi'}^2 - 2 Q \phi' - Q' + P =0\,\, ,
\label{C}\\
&& Q(\eta)= \frac{d \ln{\sqrt{- g}}}{d\eta},~~~P= 2 ( 2 {\cal H}{\cal F}
+  {\cal F}{\cal G})
\label{D}
\end{eqnarray}
where the prime denotes now the derivative with respect to $\eta$ and 
where ${\cal H} = a'/a$, ${\cal F} = b'/b$, ${\cal G} = c'/c$. 
By linearly combining the previous equations and by defining 
$\Phi = \phi - \ln{c}$ the equation for $\Phi$ 
\begin{equation}
\Phi''= \frac{B^2}{2} e^{- 2 \Phi} ,
\end{equation}
can be integrated directly twice with the result that 
\begin{equation}
\Phi = \ln{\biggl[ \frac{\delta}{\gamma} \cosh{\gamma(\eta - \eta_0)}\biggr]},
\end{equation}
where $\delta = B/\sqrt{2}$; $\gamma$ and $\eta_0$ are integration constants.
By now going to Eq. (\ref{B}) and by making the ansatz 
$a(\eta) = e^{\Phi} f(\eta)$ we get an equation 
for $f(\eta)$ 
\begin{equation}
\biggl(\frac{f'}{f}\biggr)' = \lambda^2 f^2, 
\end{equation}
which, again, can be directly integrated with the result that 
\begin{equation}
f(\eta) = -\frac{\beta}{\lambda | \sinh{\beta (\eta - \eta_0)}|}
\end{equation}
Putting all our results together we obtain that 
\begin{eqnarray}
&&\phi(\eta) = \ln{c_0} + {\cal G}_0 \eta + \ln{\biggl[ 
\frac{\delta}{\gamma} \cosh{\gamma(\eta - \eta_0)}\biggr]},
\label{s1}\\
&& a(\eta) = -\frac{\beta \,\delta}{\lambda\,\gamma} 
\frac{\cosh{\gamma(\eta- \eta_0)}}{|\sinh{\beta(\eta - \eta_0)}| },
\label{s2}\\
&& c(\eta) = c_0 e^{{\cal G}_0 \eta}.
\label{s3}
\end{eqnarray}
By requiring the consistency of this solution with the constraint equation we 
find that $ 2 \beta^2 = \gamma^2 + {\cal G}_{0}^2$.

It is interesting to notice that the solution we just obtained 
can be related to the solution of the same set of Eqs. (\ref{phi})--(\ref{F}) 
in a  Kantowski-Sachs \cite{4} metric with a magnetic field 
oriented along the radial coordinate. The Kantowski-Sachs 
line element can be written as 
\begin{equation}
ds^2 = dt^2 - m^2(t) dr^2 - n^2(t)[ d\theta^2 + A^2(\theta) d\chi^2].
\label{KS}
\end{equation}
where $A(\theta) = \sin{\theta},\,\,\theta,\,\, \sinh{\theta}$.
If a magnetic 
field directed along the radial direction is present, then, the equations of 
motion (\ref{R})--(\ref{F}) will become 
\begin{eqnarray}
&&{\cal M}' = 0, \,\,
\label{ks1}\\
&& {\cal N}' = - k n^2 m^2  e^{-2 \phi} + \frac{B^2}{2} m^2 
e^{ - 2 \phi}\,\, ,
\label{ks2}\\
&&\phi'' + {\phi'}^2 - 2 Q \phi' - Q' + P =0\,\, ,
\label{ks3}\\
&& Q(\eta)= ( {\cal M} + 2{\cal N}) ,~~~P= 2 ( {\cal M} {\cal N} + 2 {\cal N}^2),
\label{ks4}
\end{eqnarray} 
where $k= +1,~0,~-1$ if, respectively,  
$A(\theta) = \sin{\theta},~\theta,~ \sinh{\theta}$. 
This system can be integrated with the same techniques we discussed 
in the Bianchi case. Notice, moreover, that in the case $k= -1$ the 
analogy with the solutions (\ref{s1})--(\ref{s3}) is complete.

We want now to discuss the cosmic-time evolution of our solutions. 
Let us focus, for instance, on the Bianchi-type II case ($\lambda = -1$).
On the basis of our results we can consistently choose 
$\beta = \gamma = 1$. The consistency relation $2 \beta^2 = \gamma^2
+ {\cal G}^2_0$  implies that ${\cal G}_0 =1$ in string units. The 
relation between $\eta$ and the cosmic time coordinate can 
be obtained by direct integration of $a^2 c e^{-\phi} d\eta = dt$. The 
result is 
\begin{equation}
\eta(t) = \ln{\bigl[ -\frac{\delta}{t} + \sqrt{1 + \frac{\delta^2}{t^2}}\bigr]}.
\end{equation}
We want, in particular, to analyze our solutions for $t< 0$.
The solutions given in Eqs. (\ref{s1})--(\ref{s3}) can then be expressed as 
\begin{eqnarray}
&& a(t) =\frac{\sqrt{t^2 + \delta^2}}{\delta},
\label{c1}\\
&& c(t) = \biggl(\frac{\delta}{t} + \sqrt{ \frac{\delta^2}{t^2} +1}\biggr)^{-1},
\label{c2}\\
&& \phi(t) = \ln{\biggl[\frac{\sqrt{ 1 + \frac{\delta^2}{t^2}}}{\frac{\delta}{t} 
+ \sqrt{ 1 + \frac{\delta^2}{t^2}}}\biggr]}.
\label{c3}
\end{eqnarray}
In order to write the solutions in this form we also have chosen $c_0 =1$ and
$\eta_0 = 0$. The choice $\eta_0 =0$ simply translates in the origin 
the curvature singularity which appears clearly in $F= \dot{c}/c$.
For $t<0$ the dilaton coupling grows. The scale factor $a(t)$ 
contracts to a minimal value determined by $\delta$, i.e.  by 
the value of the magnetic field in string units. The scale factor 
$c(t)$ expands towards a singularity at $t=0$. 

It is finally interesting to compute the shear 
parameter in the case we just described.  The shear parameter 
measures the degree of isotropization of a given anisotropic solution
\cite{5}. We define the shear parameter, in our case , as 
\begin{equation}
r(t) = \frac{ H- G} { H + 2 G}.
\end{equation}
Using eqs. (\ref{c1})--(\ref{c2}) we find that 
\begin{equation}
r(t) = \frac{\sqrt{ t^6 + \delta^2 t^4} - \delta^3 - \delta t^2}{2 
\sqrt{t^6 + \delta^2 t^4} + \delta^3 + \delta t^2}.
\end{equation}
By expanding the above expression in Taylor series around 
$t=0_{-}$ we have that 
\begin{equation}
r(t) = -1 + \frac{\delta + 2 }{\delta^2} t^2 + {\cal O}(t^3).
\end{equation}
namely we see that the share parameter grows, a behavior similar to 
the one discussed in the Bianchi-type I case\cite{1}. This behavior is then 
not surprising and it is due to the fact that, in our discussion, 
we did not take into account the string tension corrections to the 
low-energy action. If string tension corrections are (naively) included 
the shear parameter is forced to decrease.

In conclusion we discussed the possible inclusion 
of a pure (Abelian) magnetic field in the 
low-energy string effective action. We found that, in the 
absence of any torsion background, a magnetic field and a time-dependent
dilaton field can be simultaneously accommodated in Bianchi-type 
I, II, III, VI$_{-1}$, VII$_{0}$ backgrounds. As an example 
of our considerations we found new exact solutions 
of the low-energy string effective action in the Bianchi-type III, $VI_{-1}$ 
and Kantowski-Sachs cases. We also discussed the physical 
properties of the obtained solutions and we argued that 
they are not qualitatively different from the ones 
deduced in the Bianchi-type I background \cite{1}. 

\end{document}